\documentclass[twocolumn]{aastex6}

\usepackage{amsmath}
\shorttitle{Dissolution \& Atmosphere Retention in Low-Mass Low-Density Planets}
\shortauthors{Chachan and Stevenson}

\begin{document}
\title{On the Role of Dissolved Gases in the Atmosphere Retention of Low-Mass Low-Density Planets}
\author{Yayaati Chachan\altaffilmark{1,2}}
\author{David J. Stevenson\altaffilmark{2}}
\affil{$^1$St John's College, Cambridge, CB2 1TP, UK; ychachan@caltech.edu \\
	$^2$Division of Geological and Planetary Sciences, California Institute of Technology, Pasadena, California, 91125, USA}

\received{2017 August 5}
\revised{2017 November 21}
\accepted{2017 December 26}
\published{2018 February 7}

\begin{abstract}
Low-mass low-density planets discovered by \textit{Kepler} in the super-Earth mass regime typically have large radii for their inferred masses, implying the presence of H$_2$--€"He atmospheres. These planets are vulnerable to atmospheric mass loss due to heating by the parent star'€™s XUV flux. Models coupling atmospheric mass loss with thermal evolution predicted a bimodal distribution of planetary radii, which has gained observational support. However, a key component that has been ignored in previous studies is the dissolution of these gases into the molten core of rock and iron that constitute most of their mass. Such planets have high temperatures ($>$2000 K) and pressures ($\sim$kbars) at the core-envelope boundary, ensuring a molten surface and a subsurface reservoir of hydrogen that can be 5-10 times larger than the atmosphere. This study bridges this gap by coupling the thermal evolution of the planet and the mass loss of the atmosphere with the thermodynamic equilibrium between the dissolved H$_2$ and the atmospheric H$_2$ (Henry'€™s law). Dissolution in the interior allows a planet to build a larger hydrogen repository during the planet formation stage. We show that the dissolved hydrogen outgasses to buffer atmospheric mass loss. The slow cooling of the planet also leads to outgassing because solubility decreases with decreasing temperature. Dissolution of hydrogen in the interior therefore increases the atmosphere retention ability of super-Earths. The study highlights the importance of including the temperature- and pressure-dependent solubility of gases in magma oceans and coupling outgassing to planetary evolution models.
\end{abstract}

\keywords{ planets and satellites: atmospheres, planets and satellites: composition, planets and satellites: formation, planets and satellites: interiors, planets and satellites: physical evolution}

\section{Introduction} \label{sec:intro}
The \textit{Kepler} mission discovered a substantial number of planets in the super-Earth regime that have low masses and low densities (LMLD). These planets fall between Earth and Neptune in the most abundant planetary category discovered by \textit{Kepler}  \citep{Howard2012, Batalha2013,Petigura2013}. Although the composition of these planets is highly degenerate, the presence of significant amounts of volatiles, especially hydrogen--helium, is necessary to explain the low bulk densities \citep{Rogers2015}. Here, we consider one possible composition: planets made of rocky cores surrounded by large hydrogen--helium envelopes. The presence of large low-mean-molecular-weight envelopes around planetary cores presents an opportunity to study the formation and evolution of planets in this interesting intermediate regime that has no solar system analog. In particular, the loss of atmospheric mass due to photoevaporation or the initial interior heat of the planet is an important phenomenon that plays a significant role in the evolution of these planets \citep{Lopez2012, Owen2013, Ginzburg2017}.

High-energy photons from the star can drive hydrodynamic winds that lead to atmospheric mass loss \citep{Watson1981, Lammer2003}. The ability of planets to retain their atmospheres has been studied widely, both to examine specific planets and a large parameter space \citep{Valencia2010, Owen2013, Lopez2012, Lopez2014}. Various models have coupled a planet's thermal evolution with the energy-limited hydrodynamic mass loss \citep{Hubbard2007, Nettelmann2011, Lopez2012, Jin2014, Chen2016} to study the evolution of planetary radii and atmospheres. Modeling attempts explained the presence of an 'evaporation valley', i.e., an absence of low-density planets
with very high stellar insolation, via atmospheric mass loss. Various studies also predicted a bimodal distribution in planetary radii, suggesting a dip in the abundance of planets with radii $\sim 2 \; \mathrm{R_{\oplus}}$ \citep{Owen2013, Lopez2014, Jin2014}. This bimodal distribution has recently gained stronger support since the release of observational evidence by \cite{Fulton2017}.

Notwithstanding the success of these models, there is a key component that has been neglected in the previous studies. Given the large mass of the atmosphere in LMLD planets and high equilibrium temperatures for close-in planets, the temperature at the interface between the envelope and the interior would be high enough to ensure a molten surface. The presence of a molten magma ocean permits the dissolution of a significant amount of hydrogen into the convective interior. This implies that the planet must have an internal reservoir of atmospheric gases that may be comparable to or greater than the external (atmospheric) reservoir.

The presence of an internal reservoir of atmospheric gases can have a significant influence on the evolution of LMLD planets. Outgassing from the interior could buffer the atmospheric mass loss due to photoevaporation. Attempts to explain the presence of large amounts of hydrogen in atmospheres via outgassing have been made in the past. Outgassing due to chemical reactions of water with iron \citep{Elkins-Tanton} has been suggested. However, only rapid outgassing processes have been considered \citep{Valencia2010} that have disregarded the temperature and pressure equilibrium that may exist between dissolved hydrogen and atmospheric hydrogen, if a substantial amount of the latter were present.

The present study investigates the role of this reservoir of dissolved gases in the evolution of the planetary atmospheres. The dissolution of gases is coupled with standard mass loss prescriptions and thermal evolution of planets and the effects on planetary atmospheres are then studied. Crucially, the evolution depends on the temperature dependence of solubility. This effect is poorly known but undoubtedly present, and is often not discussed in the standard literature on outgassing because of the presumed small temperature range appropriate to conventional volcanic processes. It becomes a large effect when a magma ocean is very massive and undergoes substantial temperature changes over time, as is expected from the thermal evolution of LMLD planets.

Our goal here is to develop a simple model that agrees with already-published models in the limit of no dissolution. In this way, we can isolate the distinctive effects of dissolution and outgassing on the preservation or lifetime of an atmosphere.

\section{Model} \label{sec:model}
\subsection{Atmospheric and Interior Modeling}
Our atmospheric model is simple but entirely adequate for our purpose, as we are focusing on a new effect (dissolution) and not small corrections from more accurate models of the atmosphere.

The effective temperature of a super-Earth planet is determined by the combination of the absorbed starlight and the internal luminosity (the cooling of the interior, primarily). However, the latter is small for the planets of interest, typically by four or more orders of magnitude (except perhaps very early in its history), and we can accordingly approximate the effective temperature by the equilibrium temperature imposed by the star.

\begin{equation}
T_{\mathrm{eq}} \sim T_{\mathrm{eff}}
\label{eq:equil_eff}
\end{equation}

A reasonable model of a gray atmosphere can then be obtained where the temperature profile is given by \cite{Hubeny2003, Burrows&Orton}:

\begin{equation}
T^4 \approx T_{\mathrm{eff}}^4 \; + \tau \;  \frac{F_{\mathrm{int}}}{\sigma} : \tau < \tau_{\mathrm{\mathrm{tr}}}
\end{equation}
     
\begin{equation}
T \approx \; T_{\mathrm{eff}} \; \bigg(\frac{\mathrm{p}}{P_{\mathrm{tr}}}\bigg)^\Gamma : \tau > \tau_{\mathrm{\mathrm{tr}}}
\end{equation}

$\tau$ is the optical depth in the atmospheric column and $\tau_{\mathrm{tr}}$ is the optical depth at the convective-radiative boundary, given by 

\begin{equation}
\tau_{\mathrm{tr}} = \frac{\sigma \; T_{eq}^4}{F_{int}}
\label{eq:tau_tr}
\end{equation}

$P$ is pressure, $P_{\mathrm{tr}}$ is the pressure at $\tau_{\mathrm{tr}}$, and $\Gamma$ is the appropriate adiabatic coefficient, approximately 0.3 for a cosmic composition. Because the density increases with depth, optical depth for any given height in the atmosphere is adequately approximated by

\begin{equation}
\tau = \rho \; \kappa \; H
\label{eq:tau_general}
\end{equation}

where $\rho$, $\kappa$, and $H$ are the local density, opacity, and the scale height, respectively. The equation for the temperature expresses the fact that in the radiative region, the outward heat flow ($F_{\mathrm{int}}$) can be carried at a small temperature gradient; i.e., the temperature is almost isothermal, until the opacity becomes so large that the temperature increases substantially for a small percentage increase in optical depth. The temperature profile then switches to the more steep adiabatic value because of the rapid increase of opacity with pressure (typical of all hydrogen-rich atmospheres). The opacity is calculated by interpolating the solar metallicity Rosseland mean opacities given in \cite{Freedman2008} and extrapolating their results whenever necessary.

We only consider the hydrogen component of the atmosphere in our models for dissolution. Helium is less soluble in magma and could be readily incorporated into the model. The gravitational acceleration in the convective part of the atmosphere is assumed to be constant.

\subsection{Mass Loss Rate}
Our philosophy is to focus on the novel feature of our work (dissolution) and accordingly our approach to mass loss is completely standard and identical to previous work, thus enabling easy comparison. The mass loss rate due to heating by x-ray and ultraviolet (XUV) flux is estimated using the energy-limited hydrodynamic considerations and given by \cite{Watson1981,Salz2015}:

\begin{equation}
\dot{M}_{\mathrm{e-lim}} \approx \frac{\epsilon \pi F_{\mathrm{XUV}} R_{\mathrm{XUV}}^3}{G M_{\mathrm{\mathrm{p}}} K_{\mathrm{tide}}}
\label{eq:mass_loss_rate}
\end{equation}

$F_{\mathrm{XUV}}$ is the flux from the parent star in the wavelength range of $\sim$ 1\AA{} to 2000\AA{}. We primarily model the planets around Sun-like stars. The variation of $F_{\mathrm{XUV}}$ with time is modeled by results obtained in \cite{Ribas2005}.

\begin{equation}
F_{\mathrm{XUV}} = 2.97 \times 10^{-2} \; t^{-1.23} \; d^{-2} \; \mathrm{J \; s^{-1} m^{-2}}
\label{eq:xuv_flux}
\end{equation}

where $t$ is the time elapsed in Gyr, beginning at $t$ = 100 Myr, and $d$ is the planet-star distance in au. However, it is important to model the $F_{\mathrm{XUV}}$ flux before 100 Myr as mass loss during nascent stage of the planet is especially important. We do so by by considering a saturated XUV flux from the star for the first 100 Myr fixed at the value of $F_{\mathrm{XUV}}$ obtained from the above equation at $t$ = 0.1 Gyr \citep{Scalo2007}.

The efficiency of conversion of XUV flux energy to usable work, $\epsilon$, contains all of the atmospheric and atomic physical processes that occur in the radiative region. The parameter $\epsilon$ can be dependent on planet mass and radius as well as the ionizing flux \citep{Murray-Clay2009, Owen2012}. This parameter is poorly constrained and there is some associated uncertainty. Keeping in line with the simplicity of our model and adopting the more widely used efficiency parameterization, we work with a default efficiency of 0.1. The effects of changing efficiency are studied as well, but they do not affect our understanding and results qualitatively.

$R_{\mathrm{\mathrm{XUV}}}$ is the height in the atmosphere at which most of the XUV flux is absorbed. It has been found that $\tau \sim 1$ for XUV photons at $P \sim$ 10 nbar for the super-Earth regime (slightly higher than hot Jupiters because of the lower scale height) and is thus found by assuming an isothermal radiative region in the atmosphere with exponential pressure profile  \citep{Murray-Clay2009, Valencia2010}. $K_{\mathrm{tide}}$ is a dimensionless parameter, usually not much different from unity, that accounts for the fact that gas molecules only need to escape to the Hill radius to escape the influence of planet's gravity \citep{Erkaev2007}.

Given both dissolved and atmospheric presence of hydrogen, we have:

\begin{equation}
\frac{\mathrm{d}}{\mathrm{d}t} [M_{\mathrm{\mathrm{atm}}} + M_{\mathrm{dissolved}}] = \dot{M}_{\mathrm{e-lim}}
\end{equation}

Mass loss rates obtained using this model compare well with values quoted elsewhere \citep{Yelle2008} and are of the order of $10^{10}$ g s$^{-1}$ at $t$ = 0 when the $F_{\mathrm{XUV}}$ is high.
 
\subsection{Thermal Evolution}
 
We consider a very simple but sufficient model for the thermal evolution of the planet:
\begin{equation}
\label{eq:therm_evl}
L_{int} = 4 \pi R_p^2 F_{int} = - C_v M_p \frac{\mathrm{d}\bar{T_p}}{\mathrm{d}{t}}
\end{equation}

where $\bar{T_p}$ is the average temperature of the convective interior and is linearly related to the temperature at the surface of the planet by a constant that depends on the planet's mass and size, $C_v$ is the specific heat energy (taken to be 1.0 J g$^{-1}$ K$^{-1}$), $R_p$ is the radius of the photosphere, and $F_{\mathrm{int}}$ is the interior heat flux of the planet.

This model ignores radioactivity as it becomes less important compared to interior heat with increasing planetary mass. One cannot assume an isothermal core as has been done for these planets in the past because their thermal diffusion timescale is longer than the age of the universe. Instead, the interior is assumed to be convective, with a surface temperature equal to the base temperature of the atmosphere, as is appropriate for a magma ocean. The average weighted temperature of the adiabatic core is linearly proportional to the surface temperature at the planet-envelope boundary. The constant depends on the mass of the planet and can be calculated from the Gr\H{u}neisen parameter. However, we find that the essential results do not depend much on the value chosen. In our study, we assume

\begin{equation}
\label{eq:proportion}
\bar{T_p} = 2.5 \; T_{\mathrm{sur}}
\end{equation}

which is most appropriate for a mass of 4--6 Earth masses \citep{Valencia2006}. The time evolution of $F_{\mathrm{int}}$ is set by the time evolution of the entropy in the convecting atmospheric layer: A higher (lower) entropy implies a lower (higher) pressure at the top of the adiabatic region (which is always fixed at $T_{\mathrm{eff}}$) which necessarily means a lower (higher) optical depth at that location and thus (by Equation~\ref{eq:tau_tr}) a higher (lower) $F_{\mathrm{int}}$. But a decrease in $T_{\mathrm{\mathrm{sur}}}$ does not always correspond to a decrease in atmospheric entropy in the convective layer because the basal pressure can decrease as atmosphere escapes, and entropy depends on both pressure and temperature.

The energy content of the atmosphere is significantly smaller than the rocky core's and the timescale of the cooling is primarily determined by the cooling rate of the planetary core.

\subsection{Solubility of Hydrogen}
The novel and crucial part of our model concerns the solubility of hydrogen, in particular its likely dependence on the magma temperature. Provided the basal pressure of the atmosphere is still in the ideal gas regime, Henry's law will apply and the solubility will be proportional to pressure. For the usual temperatures of terrestrial magmas, we are guided by \cite{Hirschmann2012}, but for the temperature dependence we must rely on measurement of other gases presented by \cite{Paonita2005}. In general, we expect an Arrhenius form

\begin{equation}
X = A \; P_{\mathrm{sur}} \; e^{-T_o/T_{\mathrm{sur}}} 
\label{eq:solubility}
\end{equation}

based on fundamental thermodynamics (the equality of chemical potentials for hydrogen in the co-existing phases), where $X$ is the mass fraction of the hydrogen dissolved in the planet, $P_{\mathrm{\mathrm{sur}}}$ and $T_{\mathrm{\mathrm{sur}}}$ are, respectively, the partial pressure (of the gas concerned) and temperature at the rocky core-envelope boundary, and $A$ and $T_o$ are constants. The parameter $T_o$ expresses the repulsive interaction of the molecule with the magma, i.e., it is determined primarily by molecular size.

This form for the temperature dependence of the solubility of hydrogen in magma is obtained from the study of solubility of noble gases in magma (see \cite{Paonita2005} and references therein). Such a connection can be made because there are similarities in the solubility behavior of hydrogen and the noble gases. For the noble gases, solubility decreases exponentially with the atomic size of the species. Hydrogen follows this relation and is more soluble than helium in magma at low temperatures \citep{Shackelford1972}. In addition, hydrogen's solubility in magma melts with different compositions, and is in agreement with what the extrapolated behavior for noble gases would predict for its molecular radius \citep{Hirschmann2012}. Hydrogen is also chemically inert in the conditions of interest, at least in the limit of low hydrogen mole fraction.

\begin{figure}
	\centering
	\includegraphics[width=9cm]{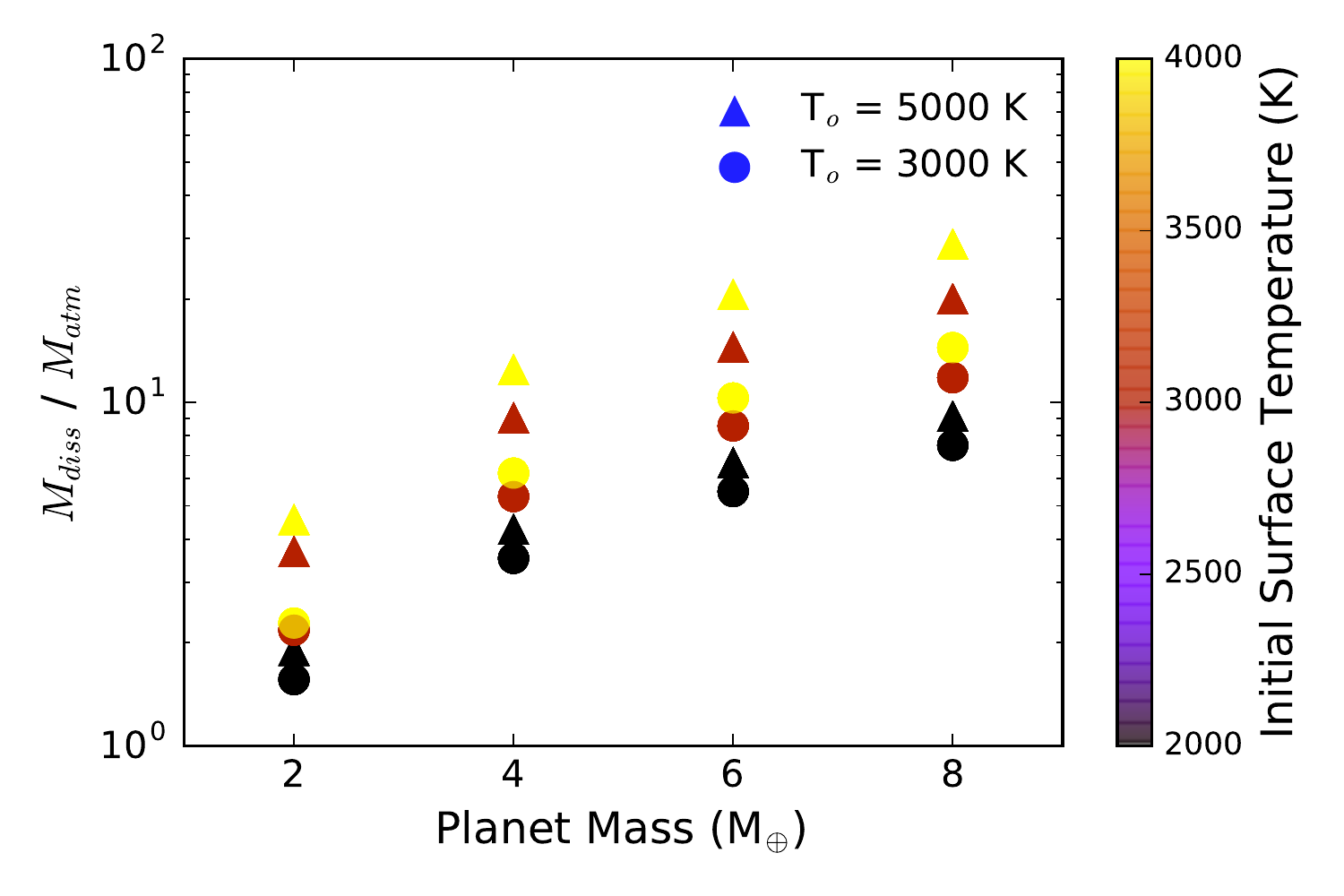}
	\caption{Ratio of amount of hydrogen dissolved in the magma to that present in the atmosphere. For a given surface temperature, a higher ratio is expected for higher-mass planets. Higher surface temperatures and $T_o$ also increase the amount of hydrogen in the interior.}
	\label{fig:m_dist}
\end{figure}

The dependence of solubility on temperature is poorly constrained by data because the petrological community is only interested in a narrow temperature range corresponding to volcanic processes on terrestrial planets such as Earth, and because of experimental uncertainties in the small range of temperature that has been studied. It is often said that solubility is only 'weakly' dependent on temperature. In fact, this is not the case if one considers the effect of doubling the temperature: the solubility can more than double. In that sense, solubility is actually more sensitive to temperature than it is to pressure. Still, the uncertainty in $T_{o}$ is quite large. Our models are evolved for different values of $T_{o}$, varying from 3000 K to 5000 K, which are consistent with the very limited data reported \citep{Paonita2005}. The constant $A$ is fixed by satisfying: $X$ = 0.001 for $P_{\mathrm{\mathrm{sur}}}$ = 1.5 kbar and $T_{\mathrm{\mathrm{sur}}}$ = 1673 K (1400$^o$ C) \citep{Hirschmann2012}. 

Figure~\ref{fig:m_dist} shows the fraction of dissolved mass to atmospheric mass and its variation with planet mass, assuming the hydrogen dissolved at the surface is the same mole fraction throughout the interior. The fraction is plotted at $t$ = 0 (before any evolution) for a variety of initial surface temperatures and two different values of T$_o$. The fraction of dissolved mass to atmospheric mass can be approximated by

\begin{equation}
\begin{split}
\frac{M_{\mathrm{diss}}}{M_{\mathrm{atm}}} \approx \frac{A \; G \; M_{\mathrm{p}}^2 \; e^{-T_o/T_{\mathrm{sur}}}}{4 \; \pi \; R^4 } \\
\approx 3 \; \bigg(\frac{M_{\mathrm{p}}}{4 M_{\oplus}} \bigg) \; exp \bigg(\frac{T_o}{1673} -\frac{T_o}{T_{\mathrm{sur}}}\bigg)
\end{split}
\end{equation}

\begin{figure*}[ht]
	\centering
	\includegraphics[width=17cm]{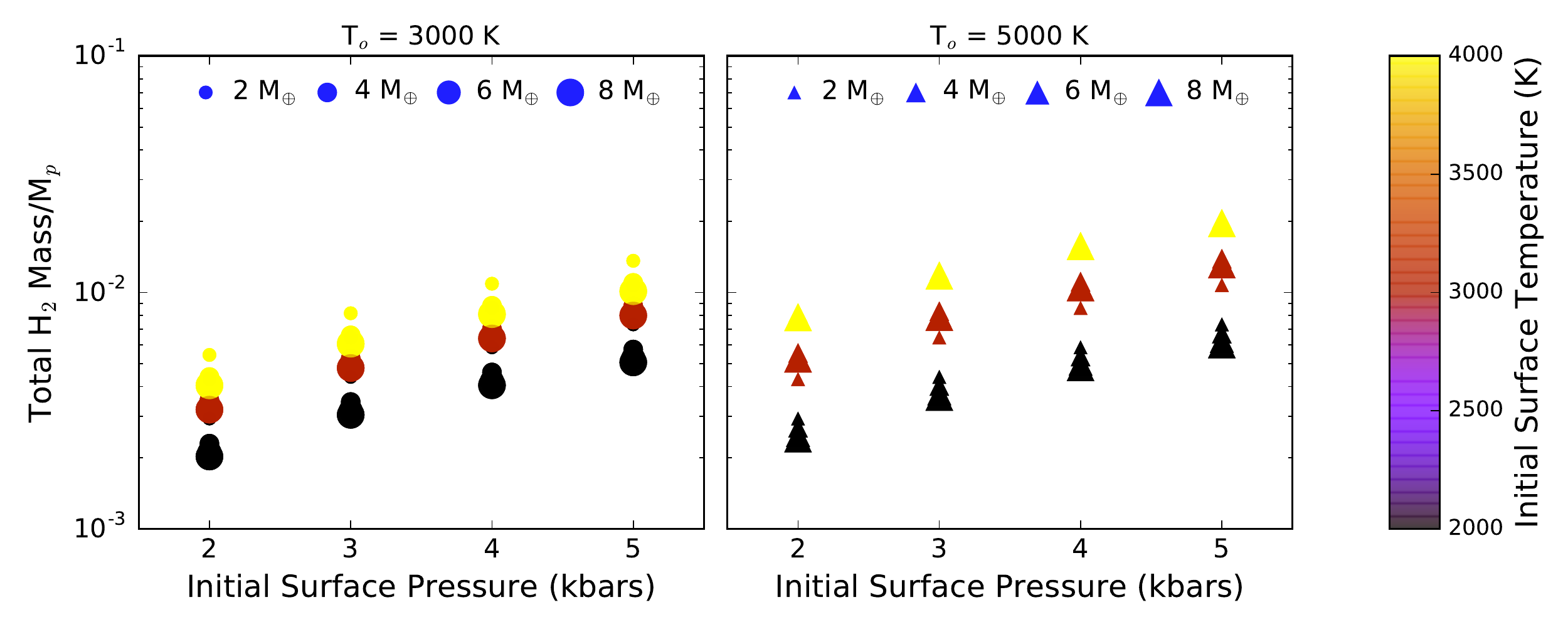}
	\caption{Variation of mass fraction of total hydrogen in our models. The size of the scatter points is indicative of planet mass. We constrain ourselves to surface temperatures $<$ 5000 K and pressures $<$10 kbar, thus obtaining  $\sim$ 1\% total mass fraction and typically 0.1\% atmospheric mass fraction. The fractional amount of hydrogen increases with both initial surface pressure and temperature.
	}
	\label{fig:sol}
\end{figure*}

where $A$ is dependent on $T_o$ , $G$ is the gravitational constant, $T_{\mathrm{\mathrm{sur}}}$ and $M_{\mathrm{\mathrm{p}}}$ are the surface temperature and the mass of the planet. The ratio is independent of surface pressure as both $M_{\mathrm{atm}}$ and $M_{\mathrm{diss}}$ are proportional to surface pressure. Figure~\ref{fig:m_dist} clearly shows the importance of including the dissolved reservoir of hydrogen in planetary evolution calculations. Depending on the surface temperature and mass of the planet, the dissolved reservoir can be 5--10 more massive than the atmospheric hydrogen content.

One thing to note, however, is that the solubility model is only valid for about $P_{\mathrm{sur}} <$ 10 kbar (ideal gas region for H$_2$) and $T_{\mathrm{sur}} <$ 5000 K. For higher temperatures, the vapor pressure of silicates would become important relative to the partial pressure of hydrogen, leading to cloud formation and considerable changes in the atmospheric temperature profile because of the latent heat effect. Even at 4000 K, there are significant effects. We have not included these effects in the first attempt to understand hydrogen storage. As previously mentioned, we also assume a convective adiabatic planetary interior that ensures the mixing and dissolution of hydrogen in the entire planetary body with a saturated solution at the surface. The solution below the surface is not saturated, as pressure and temperature rise along the adiabat in the planetary interior and the solubility increases with temperature and pressure in our prescription. In other words, the ocean never boils.

Assuming an equilibrium condition between the atmospheric hydrogen and dissolved hydrogen considerably simplifies the problem. It is possible for the thermodynamic system to be out of equilibrium. Giant impacts, which are common during early planet formation epochs, can deliver volatiles and supply the interior with a larger hydrogen content than the atmosphere \citep{Nakajima2015}. The ingassing process can be aided by impacts, but it is not guaranteed to be efficient. In the absence of impact stirring, convection may not necessarily accomplish the ingassing with high efficiency because hydrogen passes through the atmosphere-magma ocean interface by diffusion and that is limited to a thin boundary layer. It is beyond the scope of the present study to incorporate planet formation theories and the stochastic nature of initial conditions into the model. Our goal is to understand the largest effect that can arise.

\begin{figure*}[t]
	\centering
	\plottwo{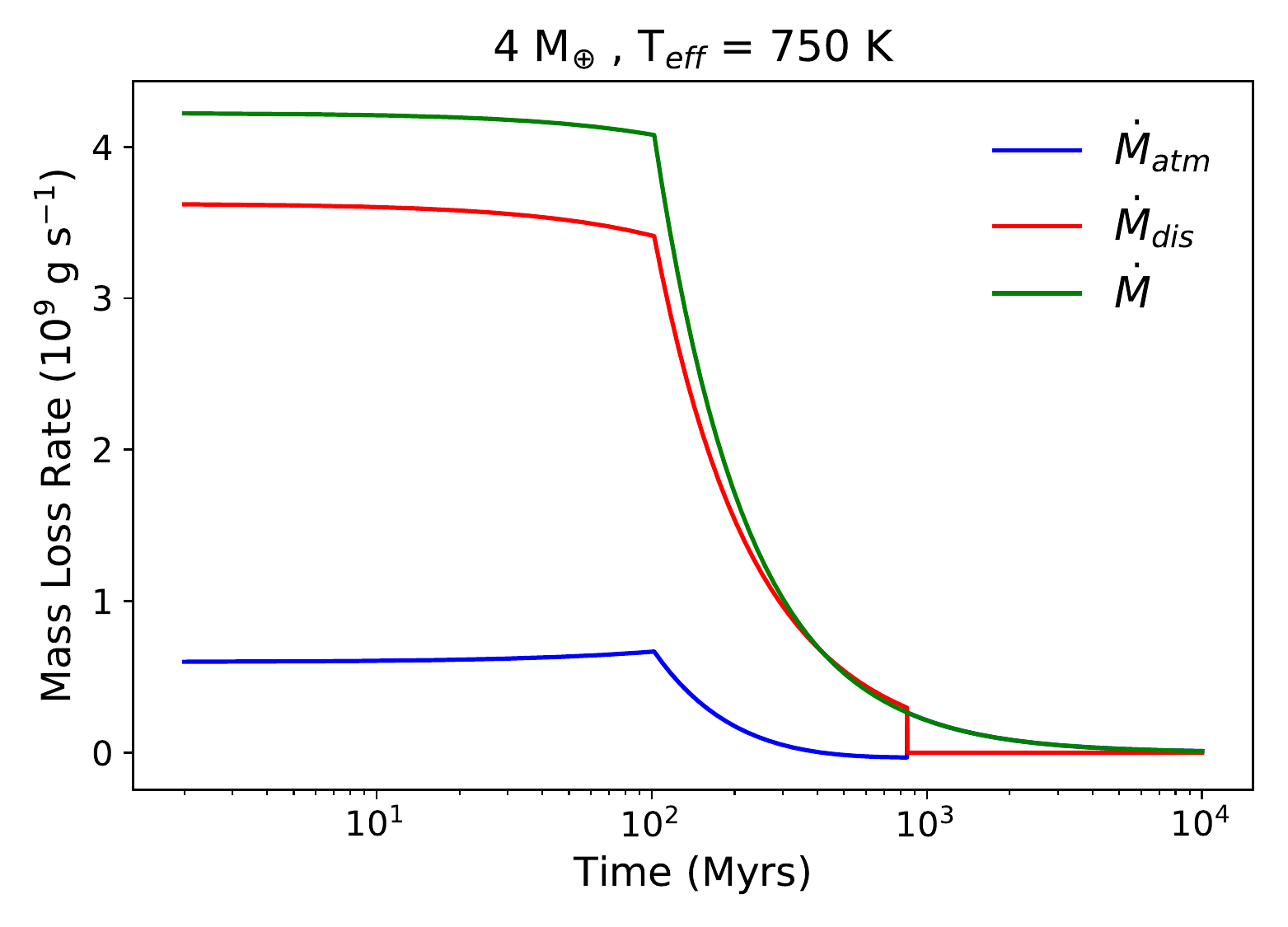}{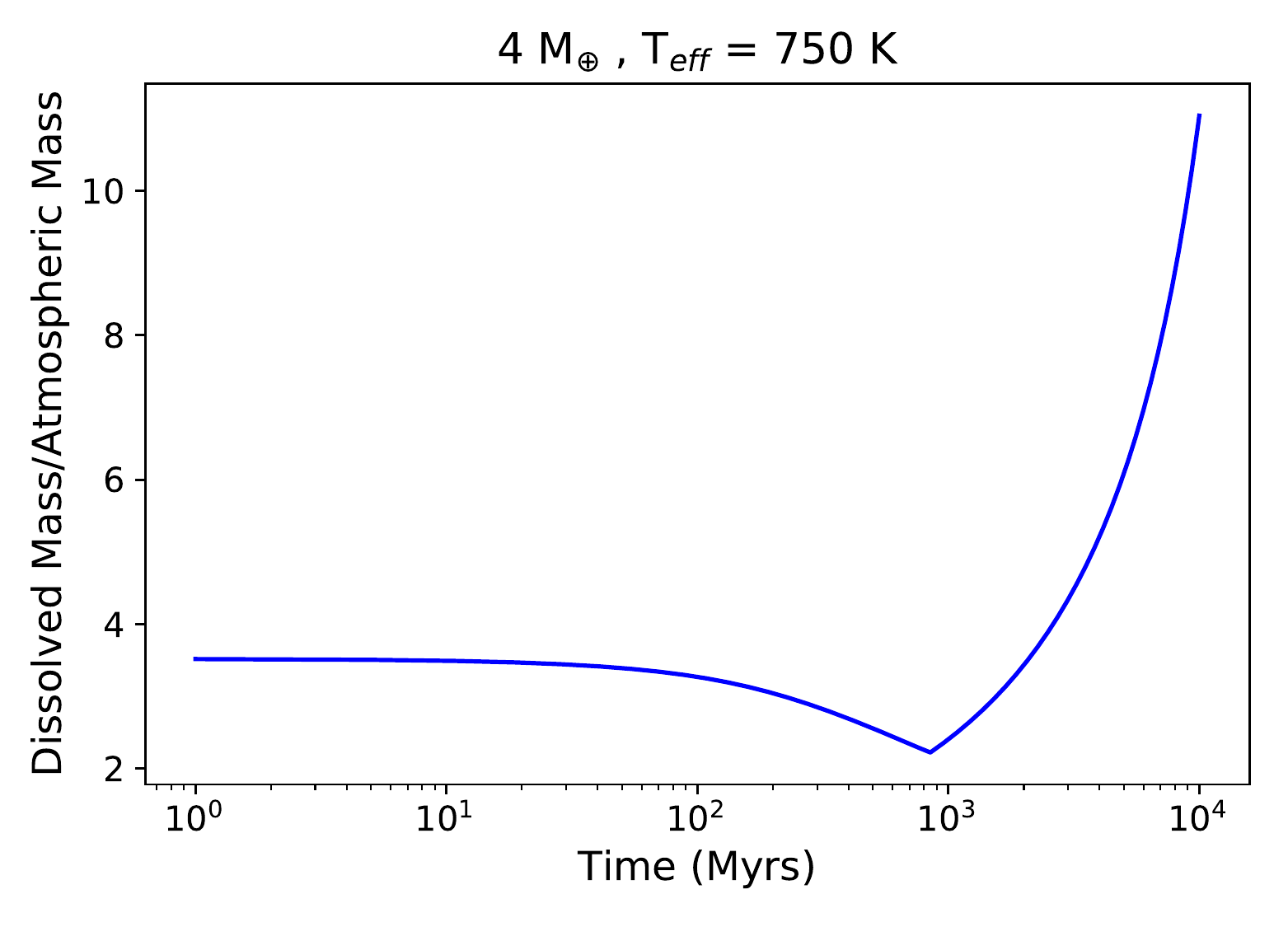}
	\caption{Evolution of a 4 M$_{\oplus}$ planet with initial T$_{\mathrm{sur}}$ = 2000 K and P$_{\mathrm{sur}}$ = 3 kbar. The overall mass loss rate for a given planet, plotted along with the mass loss rates from the atmosphere and the interior on the left. Most of the contribution to the overall mass loss is compensated for by outgassing from the interior. After 400 Myr, the outgassing from the interior is large enough to increase atmospheric mass and compensates entirely for the mass lost simultaneously. The right panel shows how the ratio of dissolved mass to atmospheric mass evolves with time. Initially, the relative mass in the atmosphere rises with time as the interior buffers the lost hydrogen. However, as the planet's surface cools to 1500 K at about 1 Gyr, the interior and the atmosphere dis-equilibrate and hydrogen is then lost entirely from the atmosphere.}
	\label{fig:m_dot_cont}
\end{figure*}

If the planet's surface cools to a temperature lower than  $\sim$1500 K, then a lid will develop atop the magma ocean and a disequilibrium between the dissolved and the atmospheric gas will start to develop. Thereafter, outgassing occurs not through a thermodynamic equilibrium process but through processes such as volcanism. We adopt a conservative approach where we do not allow any gas exchange between the planet and the atmosphere, and seal off the interface. This feature certainly involves a few simplifications. First, the formation of such a lid is of course a more gradual process than has been modeled here, but because such an event occurs only during the late times of evolutionary epochs, the exact manner of closing the lid does not have a large impact on the outcome of the model. Secondly, the formation of a lid does not totally cut off the outgassing from magma, which will occur perhaps more sporadically by volcanism. Thirdly, an important simplification is the assumption that the deep interior does not start solidifying before the top of the magma ocean does. However, this may not necessarily be true. The deep interior might start solidifying and separating from the convective magma layer first because the temperature required for that to happen is considerably higher at the pressures involved and will be reached before the surface reaches 1500 K. This would raise the hydrogen mole fraction in the magma ocean in proportion to the ratio of unsolidified magma ocean to initial magma ocean mass and thus cause additional outgassing. This can buffer further outgassing by delaying the cooling at the base of the atmosphere.

\subsection{Initial Parameters and Assumptions}
The coupled equations describing atmospheric loss, equilibrium at the base of the atmosphere, and the cooling of the planet are solved with a timestep of 1 Myr for a total duration of 10 Gyr. A grid of models is developed for the following four variables: $M_{\mathrm{\mathrm{p}}}$, $T_{\mathrm{eff}}$, $P_{\mathrm{\mathrm{sur}}}$, and $T_{\mathrm{\mathrm{sur}}}$, where the latter two quantities are defined at $t$ = 0. Other parameters such as $\epsilon$ and $T_o$ are also varied to obtain constraints on the effects of their variation.

We evolve the model for a grid of initial surface temperature and pressure because these quantities are poorly constrained by current planetary formation models. In particular, we adopt the view that giant impacts may be among the events that happen during planet formation and provide a means of ingassing, in addition to what can arise just by thermal convection and the need to establish a hydrostatic state that joins the planet atmosphere continuously to the nebula in the first few million years. One does expect a higher surface temperature and pressure for a more massive planet because of energy arguments (high gravitational potential energy) and the planet's increased ability to capture more hydrogen from the nebula before it disappears.

A key assumption is that we start the model with the hydrogen in the atmosphere equilibrated with the magma surface and uniformly mixed throughout the interior. This assumes efficient ingassing.

\section{Results}
\subsection{Solubility and Distribution of Hydrogen}

In Figure~\ref{fig:m_dist}, we show that a vast amount of hydrogen can be stored inside the planet ($\sim$ 5--10 times the atmospheric mass) for a given surface temperature and pressure. Figure~\ref{fig:sol} shows the hydrogen repository of a planet as a fraction of the total planet mass. Total hydrogen content is roughly $\sim1\%$ of the planetary mass in our models. This is because of the upper limits on temperature and pressure that are imposed by the model's assumptions. Regardless, planetary hydrogen content is sufficient to allow a study of the planet's vulnerability to mass loss and the atmosphere-interior exchange.

Because any loss from the atmosphere tends to lower the temperature and pressure at the surface, there is always a tendency for the interior to buffer the atmospheric hydrogen. This is shown explicitly in Figure~\ref{fig:m_dot_cont} (left panel), where the total loss of hydrogen is primarily balanced by the reduction of hydrogen in the interior. At around 400 Myr, the outgassing from the interior actually starts increasing atmospheric mass and compensates entirely for the mass lost simultaneously. This is the case until about 1 Gyr into the planet's evolution, when the surface temperature drops below 1500 K and the interior is sealed off from the atmosphere. Mass is then lost entirely by the atmosphere and the dissolved hydrogen stays locked inside in our models.

\begin{figure*}
	\centering
	\includegraphics[width=14cm]{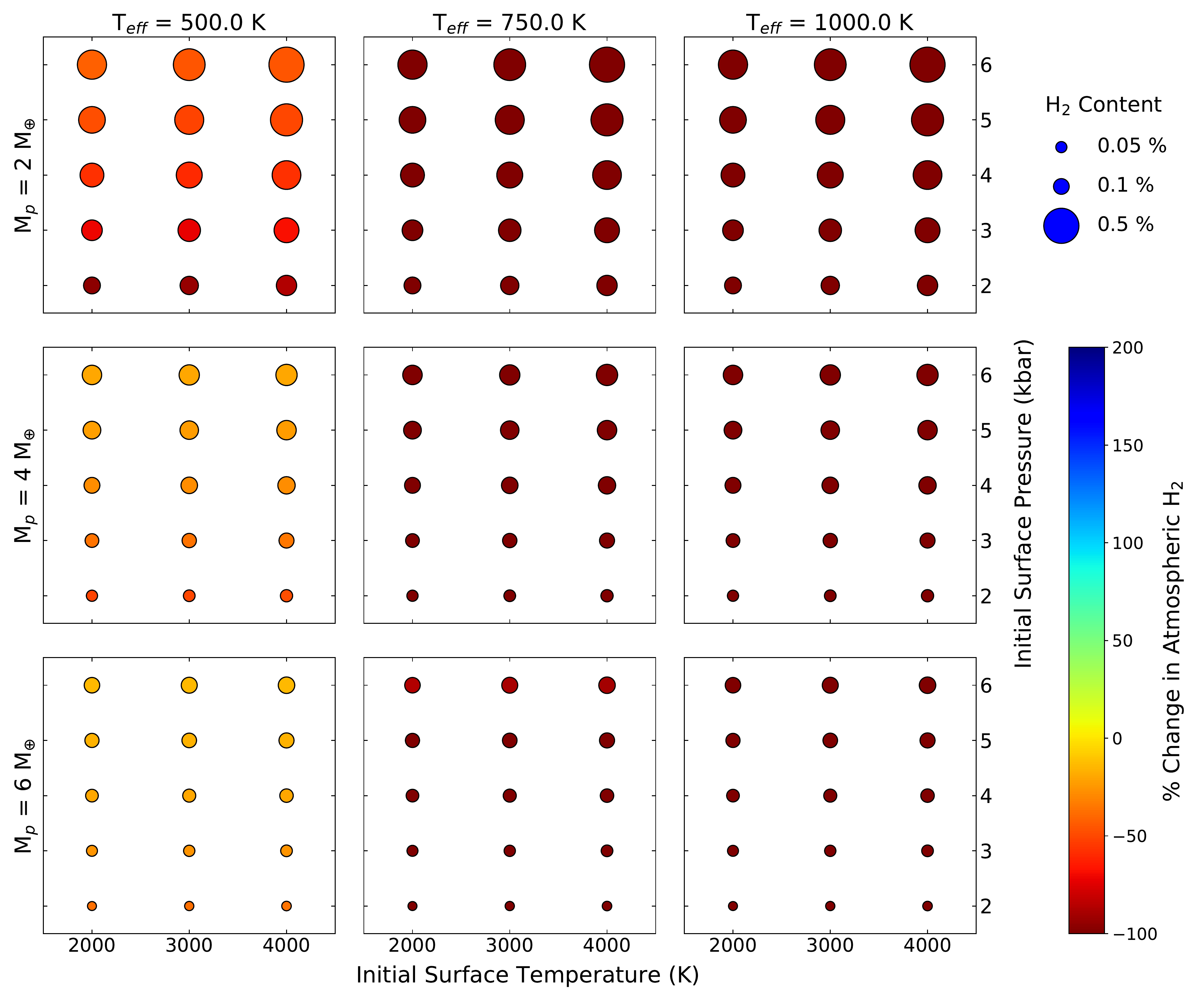}
	\caption{Plot with ratio of the final (after 10 Gyr) and initial amount of hydrogen for different planets within the given parameter space. Models are plotted on a grid of initial $T_{\mathrm{\mathrm{sur}}}$ and initial $P_{\mathrm{\mathrm{sur}}}$ in each panel. Different panels correspond to different $M_{\mathrm{p}}$ and $T_{\mathrm{eff}}$. The size of each point indicates the initial hydrogen mass fraction of the planet. The color bar indicates the relative percentages of the atmosphere retained after 10 Gyr. A planet's atmosphere retention ability increases with planet mass and initial $P_{\mathrm{\mathrm{sur}}}$, and decreases with $T_{\mathrm{eff}}$. None of the planets in the second and third column ($T_{\mathrm{eff}}$ = 750 and 1000 K) are able to retain their atmospheres.}
	\label{fig:atm_2_nd}
\end{figure*}

Figure~\ref{fig:m_dot_cont} (right panel) shows the evolution of the ratio of dissolved mass to atmospheric mass as the model is evolved for given parameters. Cooling of the planetary interior leads to outgassing from the interior, and therefore the relative mass in the atmosphere rises with time. There is an abrupt change in behavior at late stages when the planet's surface cools to 1500 K and a surface lid develops. Thereafter, the hydrogen inside the planet is preserved, while the atmospheric hydrogen continues to escape. This leads to a fractional increase of the total hydrogen dissolved inside the planet. However, escape rates are much slower at this late stage so that in some models, an atmosphere is still retained after billions of years.

\subsection{Dissolution vs Non-Dissolution}
We run a grid of models to demonstrate the impact of the dissolution of hydrogen inside the planet and the presence of such a reservoir on the atmospheric retention ability of planets. The dissolution model described above is compared with a non-dissolution model that is identical in all respects other than the solubility component. The latter serves to represent the currently adopted models in literature.

\begin{figure*}
	\centering
	\includegraphics[width=14cm]{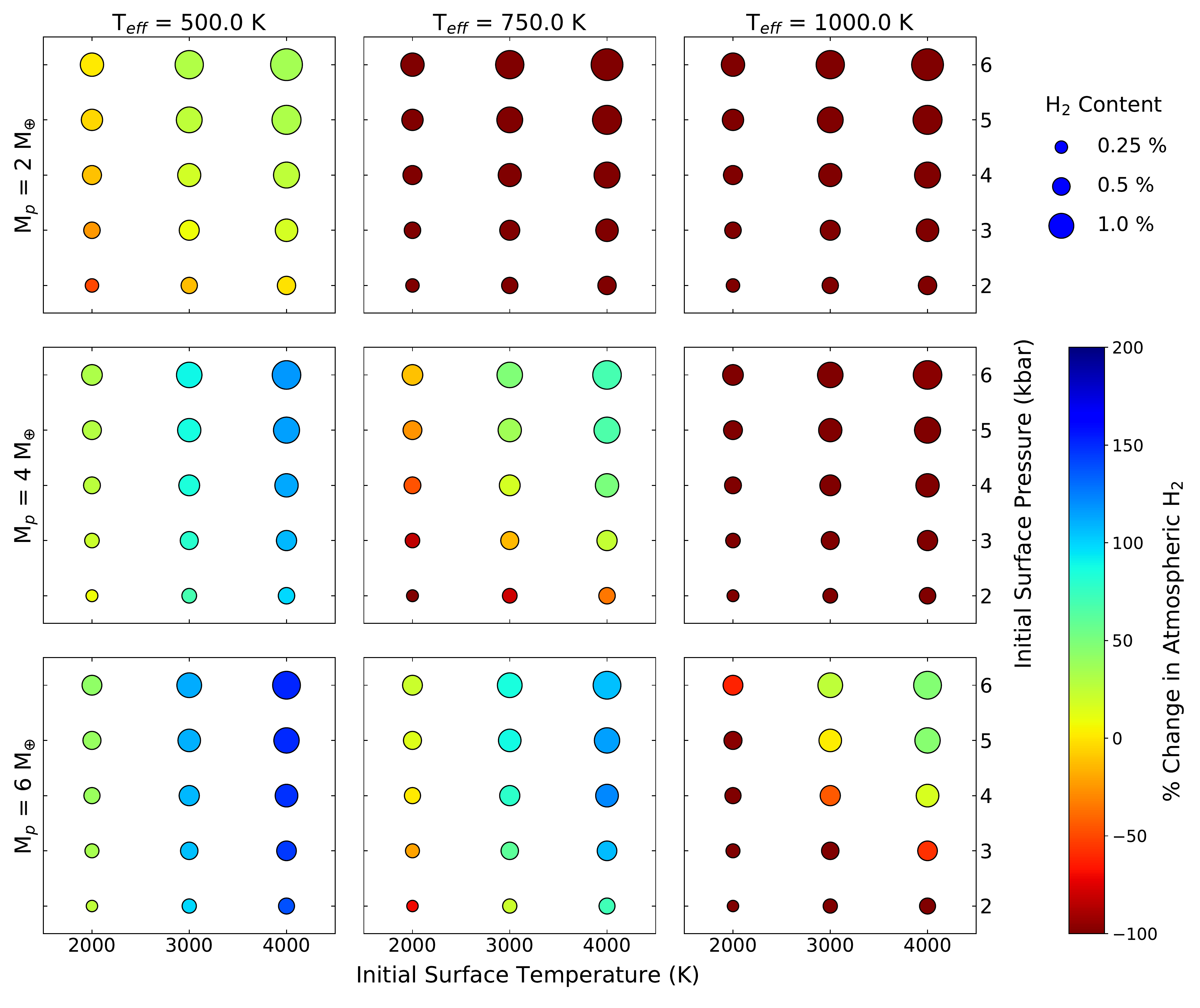}
	\caption{Plot with ratio of the final (10 Gyr) and initial amount of hydrogen for different planets within the given parameter space. Models are plotted on a grid of initial $T_{\mathrm{\mathrm{sur}}}$ and initial $P_{\mathrm{\mathrm{sur}}}$ in each panel. Different panels correspond to different $M_{\mathrm{p}}$ and $T_{\mathrm{eff}}$. The size of each point indicates the initial total hydrogen mass fraction of the planet. Note that the point size scale is different from Figure~\ref{fig:atm_2_nd} here as planets acquire more hydrogen from the nebula in the dissolution model. The color bar indicates the relative percentages of the atmosphere retained after 10 Gyr. This plot contrasts greatly with the preceding plot obtained for non-dissolution model. A number of planets in the same parameter space that were previously completely stripped of their atmospheres are able to retain them even after 10 Gyr when dissolution is taken into account. In fact, some planets' atmospheric mass increases after evolution because of continued outgassing from the interior as the planet cools and the solubility goes down. Atmosphere retention ability in this model increases with initial $T_{\mathrm{\mathrm{sur}}}$ because a higher $T_{\mathrm{\mathrm{sur}}}$ implies a larger internal hydrogen reservoir.}
	\label{fig:atm_2_d}
\end{figure*}

Because the dissolution model has the added ability to store hydrogen in the planetary interior, sensible initial conditions need to be considered in order to draw a meaningful comparison. In planet formation models, before dissipation of gas in the disk, a planet's atmosphere is assumed to be in equilibrium with the nebula at a specific radius, $R_{\mathrm{Hill}}$ or $R_{\mathrm{Bondi}}$ \citep{Ikoma2012}. If the planet has the ability to dissolve hydrogen inside its interior, there will be an active thermodynamic system at the interior-envelope boundary (R$_{\mathrm{surface}}$) as well. Equilibrium between the dissolved and the atmospheric hydrogen at the surface will co-exist with the equilibrium at the Bondi or Hill radius. An active exchange between the interior and the atmosphere therefore implies that the planet can acquire more hydrogen from the nebula. The initial condition that matches this expectation is equating the atmospheric hydrogen content in the dissolution and the non-dissolution model. This is equivalent to equating the initial surface pressure and temperature on a planet in both models.

Figure~\ref{fig:atm_2_nd} \& \ref{fig:atm_2_d} show the result of the model evolution on the planet's atmosphere over a duration of 10 Gyr for both the dissolution and the non-dissolution model. The plots shows the grid of $M_{\mathrm{\mathrm{p}}}$, $T_{\mathrm{eff}}$, $P_{\mathrm{\mathrm{sur}}}$, and $T_{\mathrm{\mathrm{sur}}}$ that was a subset of our parameter space. The color indicates the \% change in the amount of hydrogen in the atmosphere. For non-dissolution models, one would only consider a color scale from 0 to -100, as atmosphere can either remain unaffected or depreciate. However, the scale has been extended to +200\% to make a direct visual correspondence with the dissolution model, where the atmosphere can actually grow with time.

It is clear that dissolution allows for super-Earths to retain their atmospheres much closer to their parent stars. For lower-mass planets, dissolution prevents the complete stripping of the planetary atmosphere. In addition, for higher-mass super-Earths, the amount of hydrogen in the atmosphere can increase after 10 Gyr because as the planet's temperature falls, the solubility of hydrogen decreases, and the rate of outgassing overtakes the mass loss rate. These results are key to the ability of dissolution models to enable planets to retain a larger fraction of their atmosphere after 10 Gyr.

The non-dissolution models that tend to retain their atmospheres for a given planet mass lie in the high-pressure regime and there is almost no dependence on temperature, despite the fact that a higher surface temperature would increase the initial hydrogen in the atmosphere. This is because a higher initial surface temperature also implies a higher $R_{\mathrm{\mathrm{XUV}}}$, thereby increasing the mass loss rate. By contrast, the dissolution models tend to retain the most hydrogen in the high-pressure high-temperature regime. Higher temperature and pressure mean a larger reservoir of hydrogen inside the planet, which overrules the higher mass loss rate that would be obtained for higher temperatures.

\begin{figure}
	\centering
	\includegraphics[width=7.2cm]{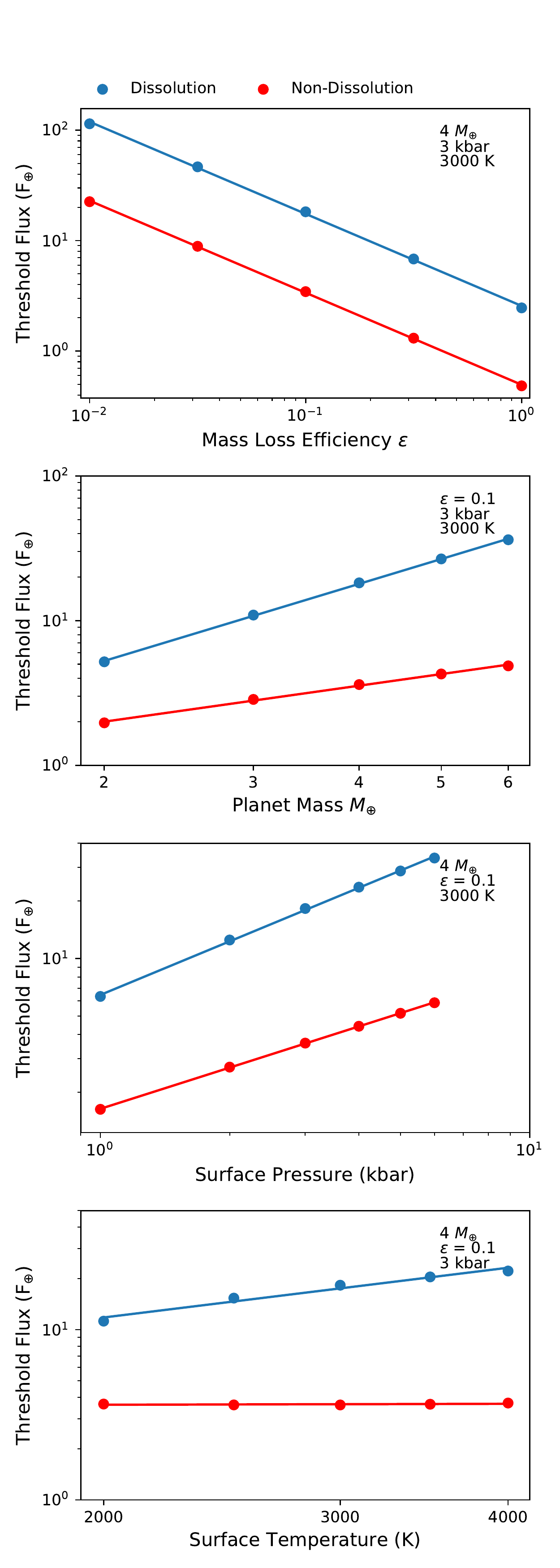}
	\caption{Power-law fitting for the threshold flux (defined as the incident flux for which only half the atmosphere is retained after 5 Gyr) against various parameters. One finds that $F_{\mathrm{th}}$ is universally higher for dissolution models with the variation of all the different parameters. In addition, the slopes for the fit against planet mass and surface temperature clearly reveal that accounting for the dissolution makes a planet more resilient to mass loss.}
	\label{fig:F_th_pl}
\end{figure}

Following  \cite{Lopez2013}, we calculate the threshold flux, $F_{\mathrm{th}}$, i.e., the flux at which the planet loses half of its atmospheric hydrogen repository in 5 Gyr. We found that the dependence of $F_{\mathrm{th}}$ on various parameters can be well described by a power-law dependence on the planet's mass (just as \cite{Lopez2013} did), the efficiency parameter in mass loss, and the surface pressure and temperature (Figure~\ref{fig:F_th_pl}). Figure~\ref{fig:F_th_pl} illustrates the power-law dependence of the threshold flux on the system parameters considered in this study. We find that the comparison of power-law indices is very informative and indicative of the vast impact that the inclusion of dissolution has on LMLD planets. The following equations yield the threshold flux for a given set of parameters:

\begin{equation}
\begin{split}
F_{\mathrm{th\_D}} \approx 18 \; F_{\oplus} \; \bigg(\frac{\epsilon}{0.1}\bigg)^{-0.8} \; \bigg(\frac{M_{\mathrm{p}}}{4 M_{\oplus}}\bigg)^{1.8} \\ 
\bigg(\frac{P_{\mathrm{sur}}}{3 \; \mathrm{kbar}}\bigg)^{0.9} \; \bigg(\frac{T_{\mathrm{sur}}}{3000 \; \mathrm{K}}\bigg)^{1}
\end{split}
\label{eq:f_th_d}
\end{equation}

\begin{equation}
\begin{split}
F_{\mathrm{th\_ND}} \approx 3 \; F_{\oplus} \; \bigg(\frac{\epsilon}{0.1}\bigg)^{-0.8} \; \bigg(\frac{M_{\mathrm{p}}}{4 M_{\oplus}}\bigg)^{0.8} \\ 
\bigg(\frac{P_{\mathrm{sur}}}{3 \; \mathrm{kbar}}\bigg)^{0.7}
\end{split}
\label{eq:f_th_nd}
\end{equation}

As expected, the relationships reveal that for an identical set of parameters, the threshold flux for a dissolution model is almost an order of magnitude higher than that of a non-dissolution model. The scaling relations clearly reveal the difference in the ability of a given planet to retain its atmosphere. The threshold flux's dependence on mass loss efficiency and initial surface pressure is similar for both dissolution and non-dissolution models, indicating slight increments in the dissolution model's performance. The difference in dependence on temperature agrees with the observation made earlier in this section. In non-dissolution models, the surface temperature's role in determining the fate of the atmosphere is weak because both the atmospheric mass and XUV radius increase with temperature, canceling out each other's effect. On the other hand, although increased temperatures imply a more extended atmosphere for dissolution models, they also indicate the presence of a much larger hydrogen repository in the planetary interior. Therefore, the atmosphere retention ability of planets increases with temperature in the dissolution models and is nearly independent of temperature in the non-dissolution models.

However, it is the mass of the rocky planet that plays the most important role for the atmosphere retention ability of these planets. The power-law exponent for planet mass in dissolution models is larger by an additive factor of 1 compared to that for non-dissolution models. The increased dependence of the threshold flux on planet mass arises because the amount of dissolved hydrogen is proportional to $M_{\mathrm{\mathrm{p}}}$. The exact power-law exponents derived here could vary for different opacity prescriptions or system parameters. However, the essence of the implications hold and these differences bear testimony to the importance of modeling dissolution of atmospheric gases in the interiors of planets.

\section{Discussion}

\subsection{Mass Loss History, Formation History, and Composition Reconstruction}
Many \textit{Kepler} planets in the low-mass low-density regime have been discovered and various studies have coupled mass and thermal evolution models to explain these worlds by invoking H$_2$--He atmospheres. Such models have revealed that these atmospheres are highly vulnerable to mass loss and require vast amounts of atmospheric hydrogen at the beginning of evolution to explain the current radius and (inferred) composition of the planet \citep{Lopez2012,Lopez2013}. However, such conclusions can be dramatically altered by the introduction of the dissolution component from our model. Accurate reconstructions of mass loss history and primordial composition are paramount to understanding the nature and formation of LMLD planets. It is therefore crucial not to neglect the dissolution of atmospheric gases inside the planet.

One of the key differences in this model and the previous models is the estimation of the amount of hydrogen available to the planet. This difference in the inferred hydrogen repository has a dramatic impact on the atmospheric longevity. The dissolution model delays the vulnerable stage of the atmosphere to a later time. The XUV flux of the star follows a power law (Equation~\ref{eq:mass_loss_rate}) and falls sufficiently by this time that the atmosphere is no longer susceptible to mass loss.

Furthermore, accounting for dissolution may also be able to explain the abundance of low-mass low-density planets at intermediate stellar fluxes and the lack of an 'evaporation valley' in the corresponding part of the parameter space. The presence of an 'evaporation valley' at high stellar insolation has substantial observational support \citep{Lundkvist2016, Fulton2017}. \cite{Lundkvist2016} reported the absence of planets with radii in the range 2.2--3.8 R$_{\oplus}$ and incident fluxes $>$650 F$_{\oplus}$. However, there are a large number of planets in Å the given radii range with incident fluxes just below the 650 F$_{\oplus}$ threshold (Figure 2 in \cite{Lundkvist2016}). An interior hydrogen reservoir could allow planets in this regime to retain their envelopes. In fact, intermediate stellar fluxes are a sweet spot for our model: the planet's distance from the star ensures that it is hot enough to retain a substantial dissolved reservoir, and not cold enough to lock the interior repository and lose its entire atmosphere.

Our models also lend some support to in situ planet formation theories. We show that if the planet remains hot, more hydrogen can be acquired from the nebula and stored in the interior of the planet. Dissolved hydrogen can then replenish the atmosphere as mass is lost. However, if the planet is further away and cools significantly before migrating inwards, the atmosphere and the interior dis-equilibrate early, a significant amount of the atmosphere is then lost, and the planet may lose its entire atmosphere. Even in this circumstance, the hydrogen in the interior may then get outgassed via volcanic processes and the planet may reacquire an atmosphere in its later stages.

\subsection{Model Uncertainties}
The model assumed fixed values for various parameters, such as the constant of proportionality, connecting surface temperature to average interior temperature, $T_{o}$, and the efficiency in mass loss estimates. These parameters were varied to evaluate whether their variation had any drastic effect on the outcomes of the model, which changed the qualitative nature of the results or the conclusions drawn thus far.

The constant of proportionality associating the surface temperature to the average interior temperature in Equation~\ref{eq:proportion}, which is assumed to be 2.5, was varied in the range 2--4 (expected range from \cite{Valencia2006}). When it was increased, the rate of change of temperature at the surface of the planet was slightly reduced, as expected from Equation~\ref{eq:therm_evl}. This led to a slower cooling of the planet and higher dissolved atmospheric mass ratio at the end of 10 Gyr. However, changing the constant did not enable any of the planets that lost their atmospheres in the initial model to retain them after 10 Gyr.

The variation of $T_{o}$, of course, has a greater quantitative effect as it significantly affects the mass fraction of hydrogen dissolved in the interior. The value of 3000 K assumed here is a conservative estimate, and $T_{o}$ was increased to 5000 K to study its influence on results obtained thus far. As one would expect, increasing $T_{o}$ led to a much larger interior reservoir of hydrogen. As a result, the lifetime ratios of the dissolved to non-dissolved models were significantly enhanced, and many more planets did not lose their atmospheres even after 10 Gyr. Increasing $T_{o}$ therefore only increases the difference between the two models.

Changing the form adopted for the opacity can have a quantitative impact on planetary evolution in our two models. We have chosen to work with opacity for a solar metallicity gas \citep{Freedman2008} to demonstrate our physical reasoning behind the phenomenon of dissolution. Increasing the metallicity and opacity actually widens the difference between dissolution and non-dissolution models. It slows down the cooling of the planet and sustains an equilibrium between the interior and the atmosphere for a longer duration, thus allowing the interior to buffer atmospheric escape late into the planet's evolution. Therefore, the quantitative results depend on the opacity prescription, but the results presented here apply generally and hold true qualitatively.

The dissolution and non-dissolution models overlap in all regards except the solubility component. This shows that the difference emerges not from some careful choice of parameters, but from a fundamental difference in the physical evolution of the planet.

\section{Conclusion}
The aim of our study is to emphasize the importance of including dissolution of atmospheric gases in surface magma oceans in planetary evolution models, especially in the context of LMLD planets. We demonstrate and explain how the dissolution model planets acquire more hydrogen from the nebula, most of which is stored in the interior. This leads to a dramatic enhancement in the ability of the planets to retain their atmospheres over a timescale of Gyr. Many more planets in the parameter space do not lose their entire atmosphere because the mass loss is compensated for by outgassing from the planetary interior. The planets also outgas over time due to their thermal evolution, which tends to increase the ratio of hydrogen in the atmosphere to that in the magma. Given the large impact including dissolution in the planet modeling has on the fate of the atmospheres, it is very important to incorporate this component in future modeling attempts and reconstruction of mass loss histories and composition. We acknowledge, however, some large uncertainties both in the ingassing process and the solubility dependence on temperature.

We thank the referees for providing valuable feedback that helped us improve our manuscript. Y.C. acknowledges the support of St John's College, Cambridge, for funding this project as part of the Undergraduate Academic Research Proposal (UARP) and Learning \& Research Fund (LRF) programmes.

\bibliography{Dissolved_Gases_Atmosphere_Retention_LMLD}

\begin{thebibliography}{}
\expandafter\ifx\csname natexlab\endcsname\relax\def\natexlab#1{#1}\fi

\bibitem[{Batalha {et~al.}(2013)Batalha, Rowe, Bryson, Barclay, Burke,
  Caldwell, Christiansen, Mullally, Thompson, Brown, Dupree, Fabrycky, Ford,
  Fortney, Gilliland, Isaacson, Latham, Marcy, Quinn, Ragozzine, Shporer,
  Borucki, Ciardi, Gautier, Haas, Jenkins, Koch, Lissauer, Rapin, Basri, Boss,
  Buchhave, Carter, Charbonneau, Christensen-Dalsgaard, Clarke, Cochran,
  Demory, Desert, Devore, Doyle, Esquerdo, Everett, Fressin, Geary, Girouard,
  Gould, Hall, Holman, Howard, Howell, Ibrahim, Kinemuchi, Kjeldsen, Klaus, Li,
  Lucas, Meibom, Morris, Pr{\v{s}}a, Quintana, Sanderfer, Sasselov, Seader,
  Smith, Steffen, Still, Stumpe, Tarter, Tenenbaum, Torres, Twicken, Uddin,
  {Van Cleve}, Walkowicz, \& Welsh}]{Batalha2013}
Batalha, N.~M., Rowe, J.~F., Bryson, S.~T., {et~al.} 2013, The Astrophysical
  Journal Supplement Series, 204, 24

\bibitem[{Burrows \& Orton(2010)}]{Burrows&Orton}
Burrows, A., \& Orton, G. 2010, in Exoplanets, ed. S.~Seager (Tucson, AZ: Univ.
  Arizona Press), 419

\bibitem[{Chen \& Rogers(2016)}]{Chen2016}
Chen, H., \& Rogers, L.~A. 2016, The Astrophysical Journal, 831, 180

\bibitem[{Elkins-Tanton \& Seager(2008)}]{Elkins-Tanton}
Elkins-Tanton, L.~T., \& Seager, S. 2008, The Astrophysical Journal, 685, 1237

\bibitem[{Erkaev {et~al.}(2007)Erkaev, Kulikov, Lammer, Selsis, Langmayr,
  Jaritz, \& Biernat}]{Erkaev2007}
Erkaev, N.~V., Kulikov, Y.~N., Lammer, H., {et~al.} 2007, Astronomy and
  Astrophysics, 472, 329

\bibitem[{Freedman {et~al.}(2008)Freedman, Marley, \& Lodders}]{Freedman2008}
Freedman, R.~S., Marley, M.~S., \& Lodders, K. 2008, The Astrophysical Journal
  Supplement Series, 174, 504

\bibitem[{Fulton {et~al.}(2017)Fulton, Petigura, Howard, Isaacson, Marcy,
  Cargile, Hebb, Weiss, Johnson, Morton, Sinukoff, Crossfield, \&
  Hirsch}]{Fulton2017}
Fulton, B.~J., Petigura, E.~A., Howard, A.~W., {et~al.} 2017, The Astronomical
  Journal, 154, 109

\bibitem[{Ginzburg {et~al.}(2017)Ginzburg, Schlichting, \& Sari}]{Ginzburg2017}
Ginzburg, S., Schlichting, H.~E., \& Sari, R. 2017, submitted to MNRAS,
  arXiv:1708.01621

\bibitem[{Hirschmann {et~al.}(2012)Hirschmann, Withers, Ardia, \&
  Foley}]{Hirschmann2012}
Hirschmann, M.~M., Withers, A.~C., Ardia, P., \& Foley, N.~T. 2012, Earth and
  Planetary Science Letters, 345-348, 38

\bibitem[{Howard {et~al.}(2012)Howard, Marcy, Bryson, Jenkins, Rowe, Batalha,
  Borucki, Koch, Dunham, Gautier, {Van Cleve}, Cochran, Latham, Lissauer,
  Torres, Brown, Gilliland, Buchhave, Caldwell, Christensen-Dalsgaard, Ciardi,
  Fressin, Haas, Howell, Kjeldsen, Seager, Rogers, Sasselov, Steffen, Basri,
  Charbonneau, Christiansen, Clarke, Dupree, Fabrycky, Fischer, Ford, Fortney,
  Tarter, Girouard, Holman, Johnson, Klaus, Machalek, Moorhead, Morehead,
  Ragozzine, Tenenbaum, Twicken, Quinn, Isaacson, Shporer, Lucas, Walkowicz,
  Welsh, Boss, Devore, Gould, Smith, Morris, Prsa, Morton, Still, Thompson,
  Mullally, Endl, \& MacQueen}]{Howard2012}
Howard, A.~W., Marcy, G.~W., Bryson, S.~T., {et~al.} 2012, The Astrophysical
  Journal Supplement Series, 201, 15

\bibitem[{Hubbard {et~al.}(2007)Hubbard, Hattori, Burrows, Hubeny, \&
  Sudarsky}]{Hubbard2007}
Hubbard, W., Hattori, M., Burrows, A., Hubeny, I., \& Sudarsky, D. 2007,
  Icarus, 187, 358

\bibitem[{Hubeny {et~al.}(2003)Hubeny, Burrows, \& Sudarsky}]{Hubeny2003}
Hubeny, I., Burrows, A.~S., \& Sudarsky, D. 2003, The Astrophysical Journal,
  594, 1011

\bibitem[{Ikoma \& Hori(2012)}]{Ikoma2012}
Ikoma, M., \& Hori, Y. 2012, The Astrophysical Journal, 753, 66

\bibitem[{Jin {et~al.}(2014)Jin, Mordasini, Parmentier, van Boekel, Henning, \&
  Ji}]{Jin2014}
Jin, S., Mordasini, C., Parmentier, V., {et~al.} 2014, The Astrophysical
  Journal, 795, 65

\bibitem[{Lammer {et~al.}(2003)Lammer, Selsis, Ribas, Guinan, Bauer, \&
  Weiss}]{Lammer2003}
Lammer, H., Selsis, F., Ribas, I., {et~al.} 2003, The Astrophysical Journal,
  598, L121

\bibitem[{Lopez \& Fortney(2013)}]{Lopez2013}
Lopez, E.~D., \& Fortney, J.~J. 2013, The Astrophysical Journal, 776, 2

\bibitem[{Lopez \& Fortney(2014)}]{Lopez2014}
---. 2014, The Astrophysical Journal, 792, 1

\bibitem[{Lopez {et~al.}(2012)Lopez, Fortney, \& Miller}]{Lopez2012}
Lopez, E.~D., Fortney, J.~J., \& Miller, N. 2012, The Astrophysical Journal,
  761, 59

\bibitem[{Lundkvist {et~al.}(2016)Lundkvist, Kjeldsen, Albrecht, Davies, Basu,
  Huber, Justesen, Karoff, {Silva Aguirre}, {Van Eylen}, Vang, Arentoft,
  Barclay, Bedding, Campante, Chaplin, Christensen-Dalsgaard, Elsworth,
  Gilliland, Handberg, Hekker, Kawaler, Lund, Metcalfe, Miglio, Rowe, Stello,
  Tingley, \& White}]{Lundkvist2016}
Lundkvist, M.~S., Kjeldsen, H., Albrecht, S., {et~al.} 2016, Nature
  Communications, 7, 11201

\bibitem[{Murray-Clay {et~al.}(2009)Murray-Clay, Chiang, \&
  Murray}]{Murray-Clay2009}
Murray-Clay, R.~A., Chiang, E.~I., \& Murray, N. 2009, The Astrophysical
  Journal, 693, 23

\bibitem[{Nakajima \& Stevenson(2015)}]{Nakajima2015}
Nakajima, M., \& Stevenson, D.~J. 2015, Earth and Planetary Science Letters,
  427, 286

\bibitem[{Nettelmann {et~al.}(2011)Nettelmann, Fortney, Kramm, \&
  Redmer}]{Nettelmann2011}
Nettelmann, N., Fortney, J.~J., Kramm, U., \& Redmer, R. 2011, The
  Astrophysical Journal, 733, 2

\bibitem[{Owen \& Jackson(2012)}]{Owen2012}
Owen, J.~E., \& Jackson, A.~P. 2012, Monthly Notices of the Royal Astronomical
  Society, 425, 2931

\bibitem[{Owen \& Wu(2013)}]{Owen2013}
Owen, J.~E., \& Wu, Y. 2013, The Astrophysical Journal, 775, 105

\bibitem[{Paonita(2005)}]{Paonita2005}
Paonita, A. 2005, Annals of Geophysics, 48, 647

\bibitem[{Petigura {et~al.}(2013)Petigura, Marcy, \& Howard}]{Petigura2013}
Petigura, E.~A., Marcy, G.~W., \& Howard, A.~W. 2013, The Astrophysical
  Journal, 770, 69

\bibitem[{Ribas {et~al.}(2005)Ribas, Guinan, Guedel, Audard, Ribas, \&
  al.}]{Ribas2005}
Ribas, I., Guinan, E.~F., Guedel, M., {et~al.} 2005, The Astrophysical Journal,
  622, 680

\bibitem[{Rogers(2015)}]{Rogers2015}
Rogers, L.~A. 2015, The Astrophysical Journal, 801, 41

\bibitem[{Salz {et~al.}(2015)Salz, Schneider, Czesla, \& Schmitt}]{Salz2015}
Salz, M., Schneider, P.~C., Czesla, S., \& Schmitt, J. H. M.~M. 2015, Astronomy
  {\&} Astrophysics, 585, L2

\bibitem[{Scalo {et~al.}(2007)Scalo, Kaltenegger, Segura, Fridlund, Ribas,
  Kulikov, Grenfell, Rauer, Odert, Leitzinger, Selsis, Khodachenko, Eiroa,
  Kasting, \& Lammer}]{Scalo2007}
Scalo, J., Kaltenegger, L., Segura, A., {et~al.} 2007, Astrobiology, 7, 85

\bibitem[{Shackelford {et~al.}(1972)Shackelford, Studt, \&
  Fulrath}]{Shackelford1972}
Shackelford, J.~F., Studt, P.~L., \& Fulrath, R.~M. 1972, Journal of Applied
  Physics, 43, 1619

\bibitem[{Valencia {et~al.}(2010)Valencia, Ikoma, Guillot, \&
  Nettelmann}]{Valencia2010}
Valencia, D., Ikoma, M., Guillot, T., \& Nettelmann, N. 2010, Astronomy and
  Astrophysics, 516, 20

\bibitem[{Valencia {et~al.}(2006)Valencia, O'Connell, \&
  Sasselov}]{Valencia2006}
Valencia, D., O'Connell, R.~J., \& Sasselov, D. 2006, Icarus, 181, 545

\bibitem[{Watson {et~al.}(1981)Watson, Donahue, \& Walker}]{Watson1981}
Watson, A.~J., Donahue, T.~M., \& Walker, J. C.~G. 1981, Icarus, 48, 150

\bibitem[{Yelle {et~al.}(2008)Yelle, Lammer, \& Ip}]{Yelle2008}
Yelle, R., Lammer, H., \& Ip, W.~H. 2008, Space Science Reviews, 139, 437

\end{thebibliography}
\bibliographystyle{aasjournal}
%% Include this line if you are using the \added, \replaced, \deleted
%% commands to see a summary list of all changes at the end of the article.
\listofchanges

\end{document}